%% file: naface.tex
\documentclass[pre,twocolumn,showpacs,showkeys,preprintnumbers,floatfix]{revtex4-1}

\usepackage{etex}
\usepackage{ifpdf}
\usepackage{hyperref}
\usepackage{dcolumn}
\usepackage{url}
\usepackage{amsmath}
\usepackage{amscd}
\usepackage{amsfonts}
\usepackage{amssymb}
\usepackage{amsthm}
\usepackage{xfrac}
\usepackage{braket}
\usepackage{bm}   
\usepackage{bbm}
\usepackage{verbatim}
\usepackage{stmaryrd}
\usepackage{xcolor}
\usepackage{setspace}
\usepackage{diagbox}
\usepackage{tabulary}
\usepackage{mathtools}
\usepackage{mathrsfs}
\DeclarePairedDelimiter\ceil{\lceil}{\rceil}
\DeclarePairedDelimiter\floor{\lfloor}{\rfloor}

\input{cmechabbrev}

\renewcommand{\H}{\operatorname{H}}

\newcommand{\kB} { k_\text{B} }
\newcommand{\cs}{\causalstate}

\newcommand{\ms}{\meassymbol}
\newcommand{\MS}{\MeasSymbol}

\newcommand{\Abet}{\ProcessAlphabet}

\newcommand{\MaxEigBeta}  { {\widehat{\lambda}_\beta} }
\newcommand{\MaxRvecBeta} { {\widehat{\mathbf{r}}_\beta} }
\newcommand{\MaxLvecBeta} { {\widehat{\mathbf{l}}_\beta} }

\newcommand{\MetricER}{ {\hmu} }

\theoremstyle{plain}    
\theoremstyle{plain}    
\theoremstyle{plain}    
\theoremstyle{plain}    
\theoremstyle{plain}    \newtheorem{The}{Theorem}
\theoremstyle{plain}    \newtheorem*{ProThe}{Proof}
\theoremstyle{plain}    
\theoremstyle{plain}    
\theoremstyle{plain}    
\theoremstyle{plain}    
\theoremstyle{plain}    \newtheorem{Def}{Definition}
\theoremstyle{plain}    
\theoremstyle{plain}

\def\clap#1{\hbox to 0pt{\hss#1\hss}}

\begin{document}

\title{Not All Fluctuations are Created Equal:\\
\vspace{0.05in}
Spontaneous Variations in\\
Thermodynamic Function}

\author{James P. Crutchfield}
\email{chaos@ucdavis.edu}
\author{Cina Aghamohammadi}
\email{caghamohammadi@ucdavis.edu}

\affiliation{Complexity Sciences Center and Department of Physics, University of
  California at Davis, One Shields Avenue, Davis, CA 95616}

\date{\today}
\bibliographystyle{unsrt}

\begin{abstract}
Almost all processes---highly correlated, weakly correlated, or correlated not
at all---exhibit statistical fluctuations. Often physical laws, such as the
Second Law of Thermodynamics, address only typical realizations---as
highlighted by Shannon's asymptotic equipartition property and as entailed by
taking the thermodynamic limit of an infinite number of degrees of freedom.
Indeed, our interpretations of the functioning of macroscopic thermodynamic
cycles are so focused. Using a recently derived Second Law for information
processing, we show that different subsets of fluctuations lead to distinct
thermodynamic functioning in Maxwellian Demons. For example, while typical
realizations may operate as an engine---converting thermal fluctuations to
useful work---even ``nearby'' fluctuations (nontypical, but probable
realizations) behave differently, as Landauer erasers---converting available
stored energy to dissipate stored information. One concludes that ascribing a
single, unique functional modality to a thermodynamic system, especially one on
the nanoscale, is at best misleading, likely masking an array of simultaneous,
parallel thermodynamic transformations. This alters how we conceive of
cellular processes, engineering design, and evolutionary adaptation.
\end{abstract}


\preprint{Santa Fe Institute Working Paper 16-09-XXX}
\preprint{arxiv.org:1609.XXXXX [cond-mat.stat-mech]}

\maketitle 

\setstretch{1.1}

\section{Introduction}

Arguably, Szilard's Engine \cite{Szil29a} is the simplest thermodynamic
device---a controller leverages knowledge of a single molecule's position to
extract work from a single thermal reservoir. As one of the few Maxwellian
Demons \cite{Maxw88a} that can be completely analyzed \cite{Boyd14b}, it
exposes the balance between entropic costs dictated by the Second Law and
thermodynamic functionality during the operation of an information-gathering
physical system. The net work extracted exactly balances the entropic cost. As
Szilard emphasized: while his single-molecule engine was not very functional,
it was wholly consistent with the Second Law, only episodically extracting
useful work from a thermal reservoir.

Presaging Shannon's communication theory by two decades, the major contribution
was that Szilard recognized the importance of the Demon's information
acquisition and storage in resolving Maxwell's paradox \cite{Maxw88a}. The
Demon's informational manipulations had an irreducible entropic cost that
balanced any gain in work. The role of information in physics \cite{Bril62a}
has been actively debated ever since, culminating in a recent spate of
experimental tests of the physical limits of information processing
\cite{Toya10a,Lamb11a,Beru2012,Jun14a,Mada14a,Peko15a,Kosk15a,Hong16a} and the
realization that the degree of the control system's dynamical instability
determines the rate of converting thermal energy to work \cite{Boyd14b}.

Hidden in this and often unstated, but obvious once realized, Maxwellian Demons
cannot operate unless there are statistical fluctuations. Szilard's Engine
cleverly uses and skirts this issue since it contains only a single molecule
whose behaviors, by definition, are nothing but fluctuations. There is no large
ensemble over which to average. The information gleaned by the engine's control
system (Demon) is all about the ``fluctuation'' in the molecule's position. And,
that information allows the Demon to \emph{temporarily} extract energy from a
heat reservoir. In the following, we ask how fluctuations are implicated more
generally in the functioning of thermodynamic systems.

To head-off confusion, and anticipate a key theme, note that ``statistical
fluctuation'' above differs importantly from the sense used to describe
variations in mesoscopic quantities when controlling small-scale thermodynamic
systems. This latter sense is found in the recently-famous fluctuation theorem
for the probability of positive and negative entropy production $\Delta S$
during macroscopic thermodynamic
manipulations~\cite{Evan93a,Evan1994,Gall95a,Kurc1998,Croo98a,Lebo1999,Coll2005}:
\begin{align}
  \frac{\Pr(\Delta S)}{\Pr(- \Delta S)} = e^{\Delta S}
    ~.
\label{eq:FT}
\end{align}
In other words, negative entropy-production fluctuations are exponentially rare
but not impossible---a fact used to great effect to determine thermodynamic
properties of biomolecules by manipulating them between macrostates
\cite{Croo99a,Liph02a,Coll05a,Alem12a}. Critically for the future, Eq.
(\ref{eq:FT}) holds out the tantalizing possibility of designing appropriately
sophisticated control systems to harvest energy from fortuitous negative
entropy-production fluctuations.

Both kinds of fluctuation are ubiquitous, often dominating equilibrium
finite-size systems and finite and infinite nonequilibrium steady-state
systems. Differences acknowledged, there are important connections between
statistical fluctuations in microstates observed in steady state and
fluctuations in thermodynamic variables encountered during general control: For
one, they are deeply implicated in expressed thermodynamic function. Is a
system operating as an engine---converting thermal fluctuations to useful
work---or as an eraser---depleting energy reservoirs to reduce entropy---or not
functioning at all?

Here, we point out a critical fact about fluctuations: \emph{they are
``processed'' by thermodynamic systems in different ways}, all other aspects
held fixed. Specifically, we show that large-deviation theory and a new Second
Law allow us to reinterpret ``fluctuations'' information-theoretically and so
identify spontaneous variations in a system's thermodynamic functioning. We
find that, in one and the same system, different fluctuations can be
transformed thermodynamically in distinct, even contradictory ways.

And this, in turn, suggests wholly new ways to take advantage of
``fluctuations'' in both the senses just described. It hints at alternative
kinds of manipulation of small-scale systems to positive benefit. We illustrate
the general idea of spontaneous variations in thermodynamic function in an
information ratchet \cite{Boyd15a}, recently introduced as an exactly solvable
model of a functional Maxwellian Demon \cite{Mand012a} and as a simple model of
a molecular \emph{information catalyst} \cite{Varn15a}. At the end, in drawing
out the consequences, we outline how these results suggest a broadened view of
information and intrinsic computing in biological systems.

\section{Thermodynamic Functioning: When is an Engine a Refrigerator?}
\label{sec:ThermoFunc}

Szilard's Engine, as we noted, and ultimately Maxwell's Demon are not very
functional: Proper energy and entropy book-keeping during their operation shows
their net operation is consistent with the Second Law. As much energy is
dissipated by the Demon as it extracts from the heat bath \cite{Szil29a}. There
is no net benefit. What about Demons that are functional?

Recently, Maxwellian Demons have been proposed to explore plausible automated
mechanisms that do useful work by decreasing physical entropy at the expense of
positive change in a reservoir's Shannon
information~\cite{Mand012a,Mand2013,Stra2013,Bara2013,Hopp2014,Lu14a,Boyd15a,Um2015}.
In particular, Boyd \emph{et al} analyzed the thermodynamics of a closely
related class of memoryful information ratchets for which all correlations
among system components---ratchet state, input and output information
reservoirs, and thermal reservoir---can be explicitly accounted \cite{Boyd15a}.

This gave an exact, analytical treatment of the thermodynamically relevant
Shannon information change from the input information reservoir (bit string
with Shannon entropy rate $\hmu$) to an exhaust reservoir (bit string with
Shannon entropy rate $\hmu^\prime$). The result was a refined and broadly
applicable Second Law that properly accounts for the intrinsic information
processing reflected in the accumulation of temporal correlations. On the one
hand, it gives an upper bound on the maximum average work $\langle W \rangle$
extracted per cycle:
\begin{align}
\langle W \rangle & \leq \kB T \ln 2 \, (\hmu' - \hmu)
   ~,
\label{eq:SecondLawInfo}
\end{align}
where $\kB$ is Boltzmann's constant and $T$ is the environment's temperature.
On the other hand, the new Second Law bounds the energy needed to materially
drive computation---transforming input information to the output information.
That is, it lower bounds the amount $-\langle W \rangle$ of input work required
for a physical system to support a given rate of intrinsic computation
\cite{Crut88a}, interpreted as producing a more ordered output---a reduction in
reservoir Shannon entropy.

As such, it subsumes Landauer's Principle \cite{Land61a,Benn82}: erasing a bit
of information irreversibly costs $-\langle W \rangle_\text{diss} = \kB T \ln
2$ in dissipated energy: The ratchet's input has $\hmu = 1$ bit/cycle and its
output output, $\hmu' = 0$ bits/cycle. Importantly, though, it goes
substantially beyond Landauer's Principle, bounding the thermodynamic costs of
general information processing transformations---that is, of \emph{any
computational process}. Critical to our purposes, though, and a consequence of
the exact analysis, Eq. (\ref{eq:SecondLawInfo})'s information-processing
Second Law allows one to identify the Demon's thermodynamic functioning.
Depending on system parameters. It acts as an \emph{Engine}, an \emph{Eraser},
or a \emph{Dud}; see Table \ref{tab:ThermoFunc} \cite{Boyd15a}.

\begin{table*}
\begin{center}
\begin{tabular}{|p{2.8cm}||p{8.3cm}|p{1.3cm}|p{2.1cm}|}
\hline
\diagbox{Function}{Feature} & Operation & Net Work & Net Computation \\
\hline
\hline
Engine	& Extracts energy from the thermal reservoir, converts it into work
		by randomizing input information
		& $\langle W \rangle > 0$
		& $\hmu' - \hmu > 0$ \\
Eraser	& Uses external input of work to remove input information
		& $\langle W \rangle < 0$
		& $\hmu' - \hmu < 0$ \\
Dud		& Uses (wastes) stored work energy to randomize output
		& $\langle W \rangle < 0$
		& $\hmu' - \hmu > 0$ \\
\hline
\end{tabular}
\end{center}
\caption{Information ratchet thermodynamic function as determined by
  Eq. (\ref{eq:SecondLawInfo}), the informational Second Law of Thermodynamics.
  }
\label{tab:ThermoFunc}
\end{table*}

\begin{figure}[!ht]
\centering
\includegraphics[width=.7\columnwidth]{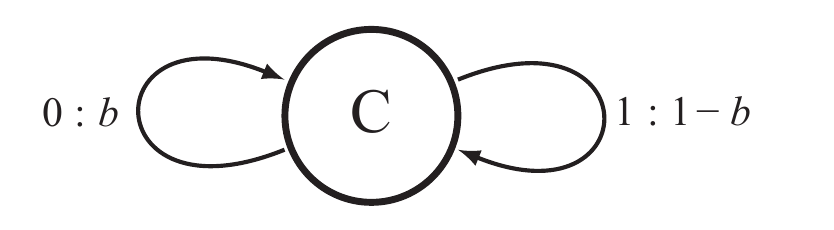} \\
(a) Input Information Reservoir\\
\includegraphics[width=.45\columnwidth]{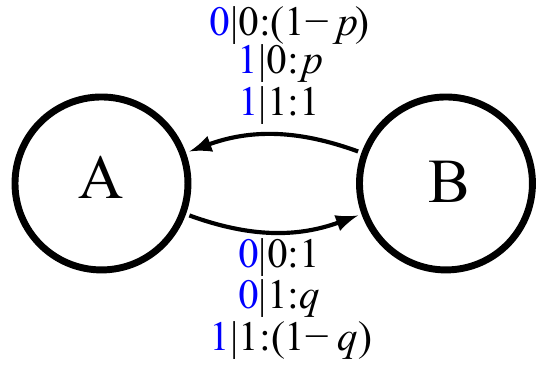} \\
(b) Input-Output Transducer\\
\includegraphics[width=.45\columnwidth]{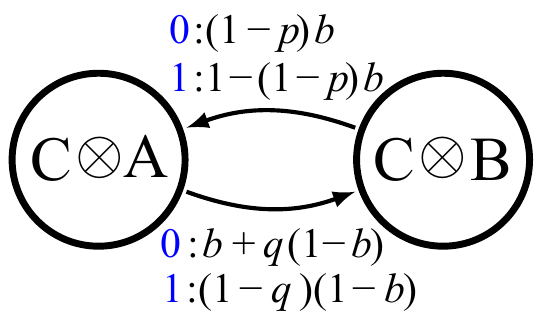} \\
(c) Output Information Reservoir
\caption{(a) Hidden Markov model that generates a biased coin input string
	$\ms_t \ms_{t+1} \ldots$ with bias $\Pr(\MS = 0) = b$. Edge
	labels $\ms : p$ indicate a state-to-state transition of
	probability $p$ that emits symbol $\ms$. (b) The information
	ratchet's transducer. Transducer edge labels
	$\ms|\ms^\prime:p$ indicate a state-to-state transition of
	probability $p$ taken on reading input symbol $\ms$ that emits
	symbol $\ms^\prime$. (c) The HMM that results from the
	transducer (b) operating on the input (a). The HMM describes
	the output string $\ms_0 \ldots \ms_{t-1}$ generated by the
	ratchet driven by a coin with bias $b$.
	(Reprinted from Ref. \protect\cite{Boyd15a} with permission.)
  }
\label{fig:Machines}
\end{figure}

To be explicit, the total work supplied by the ratchet and an input bit from a
coin of bias $b$ is
\cite{Boyd15a}:
\begin{align}
\langle  W \rangle
  & = \frac{\kB T}{2} [ (pb-q+qb) \ln{\left(\frac{q}{p}\right)} \nonumber \\
  & \quad\quad + (1-b)q \ln (1-q)+pb \ln (1-p) ]
  ~.
\label{eq:Work}
\end{align}
Here, $p$ and $q$ are parameters controlling the ratchet's detailed-balance
thermodynamics and, ultimately, its functioning. The explicit role of the
parameters is given in Figures \ref{fig:Machines}(a)-(c) which depict the
(unifilar) hidden Markov models (HMMs) for the input process, ratchet
transducer, and output process, respectively. In fact, these models are \eMs\
of the input and output processes and the \eT\ of the controller. Let's quickly
review how these process models are defined \cite{Crut12a}.

\begin{Def}
A process $\Process$'s \emph{\eM} $M(\Process)$ is the tuple $\big\{
\CausalStateSet, \{ T^{(\ms)}: \ms \in \MeasAlphabet \}, \bra{\eta_0} \big\}$,
where $\CausalStateSet$ is $\Process$'s minimal set of predictively optimal
states or causal states, $T^{(\ms)}$ are the state-to-state transition
matrices, $\MeasAlphabet$ is the alphabet of generated symbols, and
$\bra{\eta_0}$ is the initial probability distribution over the causal states.
\end{Def}

\emph{\ETs} are defined similarly, except their causal states capture how
the output process is conditioned on the input process \cite{Barn13a}.

\EMs\ are unifilar: There is at most one transition labeled with a given symbol leaving a state. A seemingly innocent syntactical property, unifilarity is
key to directly calculating $\Process$'s entropy rate $\hmu$ from its \eM\
representation $M$, as the causal-state averaged transition uncertainty:
\begin{align}
\hmu(M) \! = \! - \!\! \sum_{\cs \in \CausalStateSet} \Pr(\cs)
	\!\! \sum_{\substack{\cs^\prime \in \CausalStateSet \\ \ms \in \Abet}}
	\Pr(\cs^\prime \! , \ms|\cs) \log_2 \Pr(\cs^\prime \! ,\ms|\cs)
  ,
\label{eq:EMEntropyRate}
\end{align}
where $\Pr(\cs)$ is the asymptotic state probability calculated from the
internal-state Markov chain transition  matrix and $\Pr(\cs^\prime,\ms|\cs)$ is
the symbol-labeled transition probability $T^{(\ms)}_{\cs^\prime,\cs}$. Due our
representing the ratchet's input and output processes with their \eMs,
unifilarity allows us to exactly calculate their entropy rates, $\hmu$ and
$\hmu'$, respectively:
\begin{align}
\hmu & = \H(b) \nonumber \\
  & \equiv - b \log_2 b - (1-b) \log_2 (1-b)
  \label{eq:InputEntropyRate} \\
  \hmu' & = \frac{\H(b(1-p))}{2}+\frac{\H((1-b)(1-q))}{2}
  \label{eq:OutputEntropyRate}
   ~,
\end{align}
where $\H(b)$ is the (base $2$) binary entropy function \cite{Cove06a}.

Equations (\ref{eq:Work}), (\ref{eq:InputEntropyRate}), and
(\ref{eq:OutputEntropyRate}) explicitly give the work done $\langle W \rangle$
and information change $\hmu' - \hmu$ from input to output as a function of
input process bias ($b$) and ratchet thermal dynamics ($p$ and $q$). Thus, in
light of Eq. (\ref{eq:SecondLawInfo}) and Table \ref{tab:ThermoFunc}, we can
exactly determine the ratchet's thermodynamic function over all of the
ratchet's parameter range; see Ref. \cite[Figs. 7 and 8]{Boyd15a}.

Or, so it would seem. Let's explore what happens when there are statistical
fluctuations. Imagine that the information ratchet is implemented in a physical
substrate with a finite number of degrees of freedom, so that
fluctuations are present.

\section{Fluctuations in Steady State}
\label{sec:FlucSteadyState}

Let's first consider the ratchet's input information reservoir, by way of
introducing our general view of statistical fluctuations. Once input
fluctuations are understood, we apply the analysis to describe
its effect on ratchet functionality.

Shannon-McMillan-Breiman theory tells us that with probability close to one
sequences $\ms_{0:\ell} = \ms_0 \ldots \ms_{\ell-1}$ generated by a stochastic
process of entropy rate $\hmu$ consist of realizations whose probabilities
scale with length $\ell$ as $\Pr(\ms_{0:\ell}) \simeq 2^{- \hmu \ell}$
\footnote{This generalizes \cite{Youn93a,Agha15a} the scaling for memoryless
processes (independent, identically distributed) presented in, for example,
Ref. \cite[Ch. 3]{Cove06a}.}. Said most simply, almost all sequences are almost
equally probable. These sequences are in the so-called \emph{typical set}:
\begin{align}
  A_\epsilon^{\ell} = \lbrace
  w \in \Abet^\ell :
  2^{- \ell(\hmu+\epsilon)} \leq \Pr(w) \leq 2^{- \ell ( \hmu - \epsilon)}
  \rbrace
  ~,
\label{eq:TypicalSet}
\end{align}
where $\Abet^\ell$ is the set of length-$\ell$ words.
It can be shown that, for a given $\epsilon \ll 1$ and sufficiently large
$\ell$:
\begin{align}
\Pr (w \in A_{\epsilon}^{\ell}) \geq 1 - \epsilon
  ~.
\label{eq:PTS}
\end{align}
In other words, the probability of seeing sequences in this set is close to
one. This gives a precise and operational definition to what one means by
``typical'' behavior. In addition, as a consequence of
Eqs.~(\ref{eq:TypicalSet}) and (\ref{eq:PTS}), the typical set has
approximately $2^{\ell \hmu}$ sequences: $\left| A_\epsilon^{\ell} \right|
\simeq 2^{\hmu \ell}$. This suggests two meanings for $\hmu$: the decay
rate of probability for words $w \in A_\epsilon^{\ell}$ and the growth rate of
their number.

That said, stochastic processes do generate sequences outside their typical set.
A 60\%-40\% biased coin for a large but finite number of flips typically
produces sequences with near 60\% Heads and 40\% Tails. More precisely, for
$\ell=1000$ flips and $\epsilon=0.01$ the typical set includes sequences having
between $58.2\%$ and $61.7\%$ Heads in them. (See App.~\ref{TSBC} for the
details of such estimates.) By increasing the number of flips the percentage of
observed Heads converges to $60\%$.

At the same time, though, the process can and does generate sequences with 55\%
Heads and 45\% Tails. The occurrence of such atypical sequences are
\emph{statistical fluctuations}---any statistic calculated from them, such as a
mean, will fluctuate from trial to trial or even, when locally averaged, within
a single long realization. Importantly, the likelihood of these fluctuations is
enhanced when examining relatively short-length realizations. (We return to
this in drawing out the ultimate consequences.)

A key question, in light of these observations, is what is the range of
fluctuations for a given process? Moreover, how are fluctuations affected by
the input process' structure and memory? By way of answering these questions
and going beyond Shannon's elementary theory for memoryless processes
\cite{Shan48a} and McMillan and Breiman's focus on typical behaviors of
memoryful process \cite{McMi53a,Brei57}, Refs. \cite{Youn93a,Agha15a} show how
to calculate the entire spectrum of statistical fluctuations for structured
processes via their \eMs. We recall only the minimal necessary methods from
there, but note that they are familiar and widely used, being central to
statistical mechanics, large deviation theory in mathematical statistics
\cite{Buck90a,Touc09a}, and the thermodynamic formalism in dynamical systems
theory \cite{Ruel78,Leco07a}.

To probe fluctuations in the informational and statistical properties of the
input process $\Process$, we could simply sample its behavior. However, we are
particularly interested in behaviors outside the typical set. And, by Cramer's
theorem \cite{Demb09}, the sequence subsets of interest are exponentially rare.
That is, while we could use $M(\Process)$ to generate long realizations and
simply wait to see all of $\Process$'s statistical fluctuations, this takes an
exponentially long time or an exponentially large number of trials. To
circumvent this, we modify the process' \eM\ presentation $M(\Process)$. Let's
say that we are interested in a particular set of words outside of the typical
set; let's call this the $\beta$-set. Instead of using $M(\Process)$, as an
alternative strategy we transform $M$ to a new \eM\ $M_\beta$ that generates a
new process whose set of typical sequences is the specific fluctuation subset
of interest---the $\beta$-set---in the original process.

Thus, we consider the parameter $\beta$ as indexing $\Process$'s
\emph{fluctuation subsets} ($\beta$-sets)---sequences that all share the same
asymptotic decay rate in their probabilities; recall Eq. (\ref{eq:TypicalSet}).
At fixed $\beta$, $M_\beta$ itself generates a new process $\Process_\beta$. As
a side benefit, since $M_\beta$ is an \eM\ we can appeal to a number of tools
to efficiently calculate various informational properties directly
\cite{Crut13a}. The final step is to simply note that $M_\beta$'s information
properties are those of the fluctuation $\beta$-set in the original process
$\Process$. Let's now describe this procedure in the operational detail needed
to explore fluctuations in thermodynamic function.


To study the fluctuation subsets---the $\beta$-sets---we consider the set
$\Abet^\ell$ of all sequences of length $\ell$. The typical set is the subset
of words $w \in \Abet^\ell$ for which $\frac {-\log_2 \Prob (w)}{\ell} \approx
\hmu$. This suggests partitioning the set $\Abet^\ell$ itself into small
fluctuation $\beta$-sets that we can then study individually. To implement
this, to each sequence $w \in \Abet^\ell$ one associates an \emph{energy
density}:
\begin{align}
U^{\ell}_w = \frac {-\log_2 \Prob (w)}{\ell}
  ~,
\label{eq:UDefn}
\end{align}
mirroring the common Boltzmann weight in statistical physics: $\Prob(w) \propto
e^{-U^{\ell}_w\ell}$ \footnote{There are alternative statistics to
which one can appeal, such as superstatistics \cite{Beck03a,Hane11a}. However,
addressing this would take us too far afield at this introductory stage.}. In
our setting of structured processes, there can be forbidden sequences $w$ for
which $\Prob(w) = 0$---those with infinite energy.

Naturally, different sequences $w$ and $v$ may lead to the same energy density,
$U^\ell_w = U^\ell_v$. Realizing this, we use definition Eq. (\ref{eq:UDefn})
to partition $\Abet^\ell$ into fluctuation subsets consisting of sequences with
the same energy $U$. Energy in this statistical setting is merely a proxy for
parametrizing \emph{classes} of equal-probability-scaling sequences. In the
limit of $\ell \to \infty$ we effectively partition $\Abet^\infty$ into a
continuous family of subsets, each with a label $U$. The sequences in each
subset all share the same decay rate. Recall that we defined a $\beta-$set in a
similar manner: All the words in one of those partitions have the same decay
rate, too. In fact, $U$ and $\beta$ are simply different ways to index the same
family of partitions.

In the set of allowed energies $U^\ell = \left\{ U^\ell_w: w \in \Abet^\ell
\right\}$ energy values may appear repeatedly. Denote the count $|\{U_w^\ell=U:
U_w^\ell \in U^\ell\}|$ of length-$\ell$ sequences $w$ with equal energy $U$ by
$N(U_w^\ell=U)$. The associated sequence set is the process' thermodynamic
\emph{macrostate} at energy $U$ and we define its \emph{entropy density}:
\begin{align}
S(U) = \lim_{\ell \to \infty} \frac{\log_2 N(U_w^\ell = U)}{\ell}
  ~
\label{eq:SofU}
\end{align}
to monitor the range and likelihood of allowed sequences (or accessible
energies). This definition closely mirrors that in statistical physics, where a
macrostate's thermodynamic entropy is proportional to the logarithm of the
number of accessible microstates.

Appendix \ref{app:FlucFromEM} shows that $S(U)$ is a well behaved concave
function of $U$. From Eqs.~(\ref{eq:TypicalSet}) and (\ref{eq:UDefn}), we see
that the typical set is that subset in $\Abet^\infty$'s $U$-parametrized
partition with entropy density $S(U) = \hmu$. Let's pursue this a bit further.
Recall the two interpretations for entropy rate $\hmu$. The first was as the
decay rate of typical-set sequence probabilities. And, for an arbitrary
fluctuation subset the decay rate was interpreted as the energy density $U$.
The second interpretation was that $\hmu$ was the growth rate of the number of
sequences in the typical set. And, for an arbitrary fluctuation subset the
growth rate was the entropy density $S(U)$. This comparison gives an
alternative definition of the typical set: the only sequence subset for which
the decay rate and growth rate are equal. For all the other fluctuation sets
$S(U) < U$ and so they are rare, exponentially so.

To calculate a process' spectrum of fluctuations---how $S$ (Eq.
(\ref{eq:SofU})) depends on $U$ (Eq. (\ref{eq:UDefn})) for the sequences
outside $\Process$'s typical set---we transform its \eM\ $M$ to a new
``twisted'' \eM\ $M_\beta$ whose typical set is $\Process$'s fluctuation subset
at $\beta$ \cite{Youn93a,Agha15a}. (Appendix \ref{app:FlucFromEM} reviews the
detailed construction of $M_\beta$.) Moreover, there is a one-to-one mapping
between $\beta$ and $U$. This means that there is a well defined, invertible
function $U(\beta)$. And so, varying $\beta$ between negative infinity and
positive infinity sweeps over all the fluctuation subsets.

Operationally, using $M_\beta$ gives a direct way to calculate the
thermodynamic entropy density and energy density as a function of $\beta$:
\begin{align}
&S(U(\beta)) = \MetricER(M_\beta)
  ~,
\label{eq:EntropyDensity}
  \\
&U(\beta) = \frac{1}{\beta} ( \MetricER(M_\beta) -  \log \MaxEigBeta ) 
  ~,
\label{eq:EnergyDensity}
\end{align}
where $\MaxEigBeta$ is $M_\beta$'s transition matrix's maximal eigenvalue. And,
using the \eM\ entropy rate expression in Eq. (\ref{eq:EMEntropyRate}) gives a
similarly direct way to calculate the thermodynamic entropy density. All in
all, using $\Process$'s \eM\ leads to explicit expressions for the fluctuation
spectrum of the process it generates: the range of fluctuations (energies
$U(\beta$)) and the ``sizes'' $S(U)$ of its fluctuation subsets.

While this exposition on fluctuations may seem indirect, there is a rather
simple and geometric description of the basic shape and properties of the
fluctuation spectrum $S(U)$. First, at a given energy, $\beta$ is the slope of
$S(U)$: $\beta = \partial S(U) / \partial U$. (Appendix \ref{app:FlucFromEM}
gives the proof.) Second, a process' typical set occurs at the $U$ such that
$\beta = 1$. Third, a process' most likely sequences occur at the extreme of
$\beta \to \infty$. Since probability is associated with energy, we think of
these sequences as a process' \emph{ground states}. That is, the lowest energy
sequences are the most probable. Fourth, for an ergodic, finite-memory process
$S(U)$ is a well behaved, convex function of $U$. Fifth, and finally, the
latter implies that there is also a set of least probable or ``high energy''
sequences, found at $\beta \to -\infty$. And so, $\beta$ can be negative,
indicating the statistical analog of the physics of population inversion. We
now turn to illustrate these properties and their consequences for
thermodynamic functioning.

\section{Functional Fluctuations}
\label{sec:FuncFluc}

We are ready to bring together our identification of thermodynamic
functionality in Sec. \ref{sec:ThermoFunc}, which ultimately derived from Eq.
(\ref{eq:SecondLawInfo})'s Second Law for information processing, with
Eqs.~(\ref{eq:EntropyDensity})'s and (\ref{eq:EnergyDensity})'s analysis of
statistical fluctuations in Sec. \ref{sec:FlucSteadyState}. With the connection
made, we then go on to calculate the likelihood of functional fluctuations.

\subsection{Setting}
\label{sec:Setting}

Recall the ratchet introduced in Sec. \ref{sec:ThermoFunc}, but with its Markov
dynamic parameters $p = 0.2$ and $q = 0.6$ and with an input reservoir
generating independent and identically distributed (IID) symbol sequences of
bias $b = 0.9$. If we operate the input reservoir for a sufficiently long time,
with high probability we observe a sequence that has nearly $90\%$ $0$s in it.
Using Eqs.~(\ref{eq:Work}), (\ref{eq:InputEntropyRate}), and
(\ref{eq:OutputEntropyRate}) we see positive work $\langle W \rangle > 0$ and
positive entropy production $\hmu' - \hmu > 0$, describing the ratchet's
transforming the input process' typical sequences to the output process. Then,
by Table~\ref{tab:ThermoFunc}, the ratchet typically operates as an engine. The
work $\langle W \rangle$, too, is function of input-process typical set and the
ratchet parameters.

As we emphasized earlier, it is not always the case that the input reservoir
generates ideal typical sequences. It also generates sequences outside the
typical set. For example, given the parameters quoted, it can generate long
sequences with $70\%$ $0$s. Let's consider the case where a long atypical
sequence $w$ is generated for which $\frac {-\log_2 \Prob (w)}{\ell} = U$,
but $U \neq \hmu$. What is the functionality of ratchet in this case?

The key here is to find an alternate process that typically generates sequences
with energy density $U$ and then analyze the ratchet's response to them. As
noted above, for every fluctuation subset with energy density $U$ there is a
unique $\beta$ such that the new process' $M_\beta$ generates this fluctuation
subset typically. Using Eqs.~(\ref{eq:SofU}) and (\ref{eq:EntropyDensity}), the
entropy rate $\hmu$ of the new process is $S(U)$. With this method we can
directly calculate $\hmu$, $\hmu'$, and $\langle W \rangle$ for $M_\beta$ and,
consequently, for the particular fluctuation subset at $U$. Putting these
quantities together, we then identify the ratchet's functionality via Table
\ref{tab:ThermoFunc}.

To keep distinct properties distinct and so reduce confusion, we must emphasize
a point about notation and interpretation. The previous section introduced a
parameter $\beta$ for a given process. Despite its mathematical similarity to
the inverse temperature in statistical mechanics and historical reasons for
using that notation, for ease of understanding $\beta$ should be thought of
simply as a index of various fluctuation subsets generated by the given
process. (Technically, we can do this since $U$ indexes the fluctuation subsets
and $U(\beta)$ is monotonic in $\beta$.) Equally important, the input process
parameter $\beta$ and the output $\beta^\prime$ parameter are conceptually
distinct from the temperature $T$ of the ratchet's thermal reservoir; e.g., as
used in Eqs. (\ref{eq:SecondLawInfo}) and (\ref{eq:Work}). In short, at this
point in our analysis, none of these three variables should be conflated
notationally nor physically.

\begin{figure}
\centering
\includegraphics[width=\columnwidth]{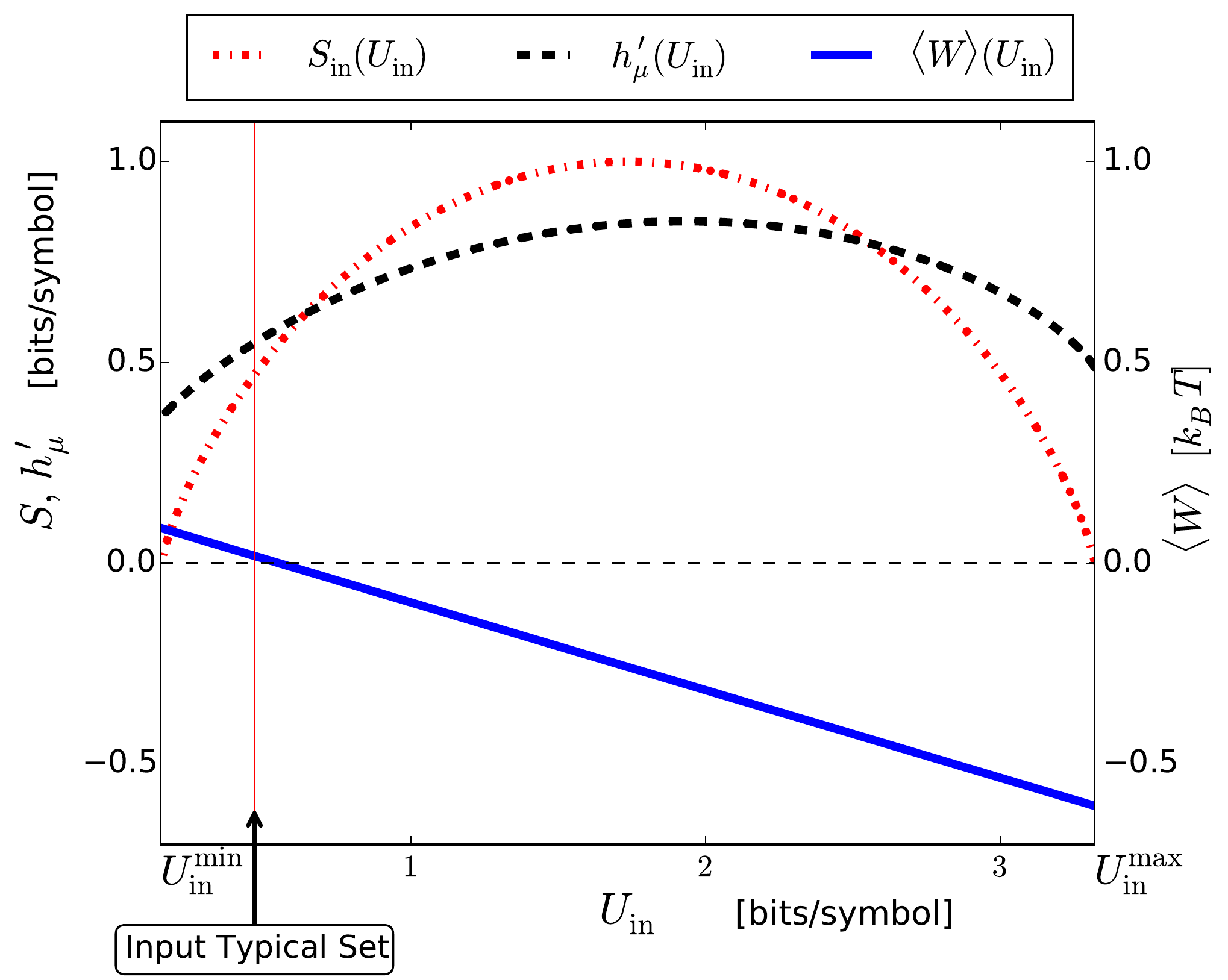}
\caption{Fluctuation spectra under a thermodynamic transformation implemented
	by Ref. \cite{Boyd15a}'s information ratchet with parameters $p=0.2$ and
	$q=0.6$ driven by an IID input source with bias $b=0.9$: Input process
	Shannon entropy rate $\hmu = S_\text{in}$ versus fluctuation-subset label
	(energy) $U_\text{in}$ (alternating dashed line), resulting output process'
	$\hmu'$ versus $U_\text{in}$ (dashed line), and the average work $\langle W
	\rangle$ versus $U_\text{in}$ (solid line).
	}
\label{fig:FlucSpecTrans}
\end{figure}

\subsection{Input, Ratchet, and Work Ratchet Fluctuations}

The first step is to determine the fluctuation spectrum for the input process
and then the spectrum of the ratchet's response. Recall that we are considering
the behavior of Ref. \cite{Boyd15a}'s information ratchet, but now as we sweep
$\beta$ we control which subsets outside the typical set we focus on and
consequently which fluctuation subset we analyze. For the analysis, recall that
the input and output processes are specified by the unifilar HMMs in Figs.
\ref{fig:Machines}(a) and \ref{fig:Machines}(c), respectively.

As $\beta$ sweeps from $-\infty$ to $\infty$, by using the new \eM\ $M_\beta$
we can analyze all of the fluctuation subsets generated by the input process.
A result of the method in App. \ref{app:FlucFromEM}, $M_\beta$ is the same as
the \eM\ in Fig.~\ref{fig:Machines}(a), except that we change $b$ to
$\widehat{b} = b^\beta / \left( b^\beta+(1-b)^\beta \right)$. The input
process' thermodynamic entropy density $S_\mathrm{in}(U_\mathrm{in})$ and
energy density $U_\mathrm{in}$ are calculated from Eqs.
(\ref{eq:EntropyDensity}) and (\ref{eq:EnergyDensity}). Then, feeding the new
process to the ratchet, $\langle W \rangle$ can be calculated from
Eq.~(\ref{eq:Work}), again by changing $b$ to $\widehat{b}$. We denote this
work quantity $\langle W \rangle(U_\mathrm{in})$. By feeding the new input
process to the ratchet the output process' \eM\ is the same as the \eM\ in
Fig.~\ref{fig:Machines}(c) but we again change $b$ to $\widehat{b}$. The
entropy rate of this output process is denoted by $\hmu'(U_\mathrm{in})$. To
predict the thermodynamic effect of feeding in the fluctuation subset with
energy density $U_\mathrm{in}$ instead of feeding it with a typical sequence,
we substitute $S_\mathrm{in}(U_\mathrm{in})$, $\hmu'(U_\mathrm{in})$, and
$\langle W \rangle(U_\mathrm{in})$ for $\hmu$, $\hmu'$, and $\langle W
\rangle$, respectively, in the informational Second Law Eq.
(\ref{eq:SecondLawInfo}).

Figure \ref{fig:FlucSpecTrans} puts these altogether, showing the input
process' fluctuation spectrum $S_\mathrm{in}(U_\mathrm{in})$, the output
process' spectrum $\hmu'(U_\mathrm{in})$, and the dissipated work $\langle W
\rangle$ versus fluctuation energy density $U_\mathrm{in}$. There are several
observations to make first, before we associate thermodynamic function.

First, let's locate the input typical set. It occurs at a $U$ such that the
slope $\beta = 1$ on $S_\text{in}(U_\mathrm{in})$. The figure identifies it
with vertical line, so labeled.

Second, the input process' ground states occur as $\beta \to \infty$. As a
consequence of Eq.~(\ref{eq:UDefn}) the ground state at
$U_\text{in}^\text{min}$ corresponds to the sequence with the highest
probability. In this case this is the all-$0$s sequence and consequently
$U_\text{in}^\text{min} = -\log_2 (b)\simeq 0.152$. The other extreme is at
$U_\text{in}^\text{max}$, corresponding to the lowest probability, allowed
sequence. In this case it is the all-$1$s sequence. Consequently,
$U_\text{in}^\text{max} = -\log_2 (1-b) \simeq 3.32$. Visualized as slopes on
$S_\text{in}$ respectively, these extremes occur at the very left and very
right portion of the curves, respectively. Note that there is only a single
sequence associated with $U_\text{in}^\text{max}$ and only one with
$U_\text{in}^\text{min}$. By using Eq.~(\ref{eq:SofU}) we have
$S_\text{in}(U_\text{in}^\text{max})= S_\text{in}(U_\text{in}^\text{min})=0$,
as seen in the figure.

Third, note that the input fluctuation spectrum $S_\text{in}(U_\text{in})$ is
rather familiar. The parametrized representation of the function
$S_\text{in}(U_\text{in})$ in terms of $\beta$ is the well known fluctuation
spectrum of a biased coin. (See App. \ref{app:WorkLinearInU}.) 

Fourth, the spectrum of output entropy rates $\hmu^\prime (U_\text{in})$ ranges
from $U_\text{in}^\text{min}$ to $U_\text{in}^\text{max}$, but does not vanish
at these extremes. This indicates stochasticity in the output process that is
added by the ratchet itself to the zero entropy-rate input sequences there.
More on the functional consequences, shortly.


Fifth and finally, to complete the task, we must determine the average work
$\langle W \rangle$ as a function of energy $U_\text{in}$. From the figure, we
see that the dissipated work $\langle W \rangle$ is linear in the energy
density $U_\text{in}$. (Appendix \ref{app:WorkLinearInU} derives this.) 

\subsection{A Spectrum of Thermodynamic Function}

So much for statistical fluctuations in the operation of the system's components individually. What does the informational Second Law tell us
about the range of thermodynamic functioning---Engine, Eraser, or Dud---the
ratchet performs when exhibiting these fluctuations? With the detailed analysis
of the input and output process fluctuation spectra and their associated
energies, we are ready to invoke the informational Second Law to determine the
ratchet's effective thermodynamic function for various fluctuations.

\begin{figure}
\centering
\includegraphics[width=1.04\columnwidth]{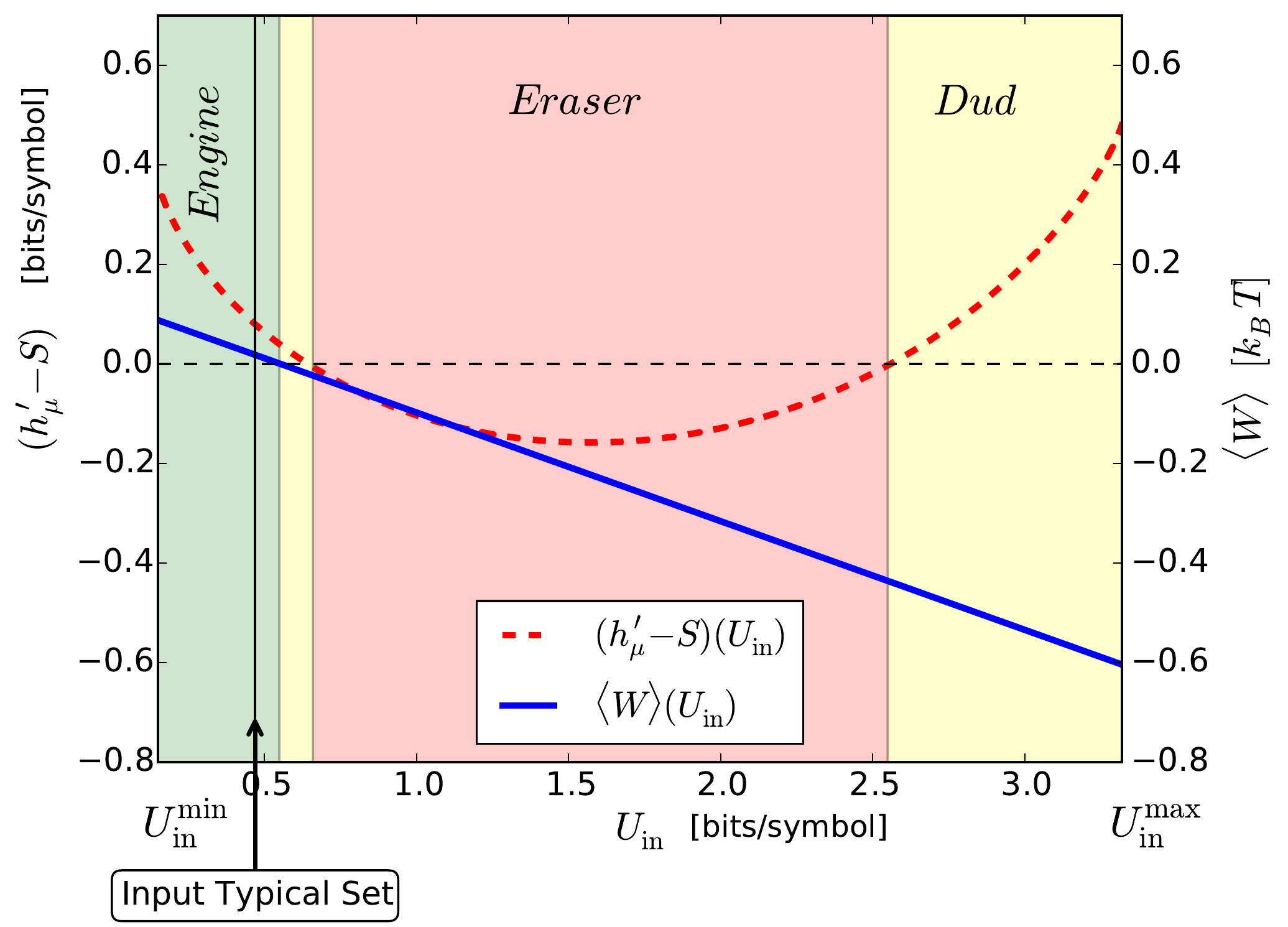}
\caption{Fluctuations in thermodynamic functioning: Using the new Second Law
  of information processing and Table \ref{tab:ThermoFunc} to identify how
  different fluctuation subsets operate within the information ratchet as a
  function of subset label $\beta \propto e^{-U}$. Engine: $\hmu^\prime
  (U_\text{in})  - S_\text{in} (U_\text{in}) > 0 $ and $\langle W \rangle > 0$;
  Eraser: $\hmu^\prime (U_\text{in}) - S_\text{in} (U_\text{in})  < 0 $ and
  $\langle W \rangle < 0$; and Dud: otherwise.  Note that the range of the
  horizontal axis extends only over the range of $U_\text{in}$, since that is
  all that can be accessed by input fluctuations; cf. Fig.
  \ref{fig:FlucSpecTrans}.
 }
\label{fig:thermofluc}
\end{figure}

Figure \ref{fig:thermofluc} summarizes this functional identification, using
the trade-offs between input-output process entropy change and dissipated work
$\langle W \rangle$ specified by Eq. (\ref{eq:SecondLawInfo}) and the
thermodynamic functioning identified in Table \ref{tab:ThermoFunc} to label the
various functional regimes parametrized by $U_\text{in}$. These fall into four
regimes, from left to right, increasing $U_\text{in}$, the ratchet operates as
an engine (green), a dud (yellow), an eraser (red), and then again as a dud
(yellow).

To better understand how the ratchet operates thermodynamically, consider the
ground state of the input process; which as just noted has only a single
member, the all-$0$ sequence with zero entropy rate $S(U_\text{in}^\text{min})
= 0$. If we feed this sequence into the ratchet, the ratchet adds stochasticity
which appears in the output sequence. The first $0$ fed to the ratchet leads to
a $0$ on the output. For the next $0$ fed-in, with probability $p$ the ratchet
outputs $1$ and with probability $1-p$ it outputs $0$. The entropy rate of
output sequence then is $\hmu'(U_\text{in}^\text{min}) = \frac{1}{2} H(p)
\simeq 0.36$.  (See also the left end of $\hmu^\prime$ in Fig.
\ref{fig:FlucSpecTrans}.)

To generate this sequence we simply use the \eM\ in Fig.~\ref{fig:Machines} with $b=1$. With this biased process as input, using Eq.~(\ref{eq:Work}) we find $\langle W \rangle(U_\text{in}^\text{min}) \simeq 0.0875 > 0$.
Table ~\ref{tab:ThermoFunc} then tells us that if we feed the ground state of
the input process to the ratchet, it functions as an engine. At the other
extreme $U_\text{in}^\text{max}$, the only fluctuation subset member is the
all-$1$s sequence with $S(U_\text{in}^\text{max})=0$. Again, the ratchet adds
stochasticity and the output has $\hmu'(U_\text{in}^\text{max}) = \frac{1}{2}
H(q) \simeq 0.485$. (See also the right end of $\hmu^\prime$ in Fig.
\ref{fig:FlucSpecTrans}.) To generate this input sequence we simply use the
\eM\ in Fig.~\ref{fig:Machines} with $b=0$. With this process as an input, we
use Eq.~(\ref{eq:Work}) again and find negative work $\langle W
\rangle(U_\text{in}^\text{max})\simeq-0.6$. Table ~\ref{tab:ThermoFunc} now
tells us that feeding in this extreme sequence (input fluctuation) the ratchet
functions as a dud.

We conclude that the ratchet's thermodynamic functioning depends substantially
on fluctuations and so will itself fluctuate over time. The Engine
functionality occurs only at relatively low input fluctuation energies, seen on
Fig. \ref{fig:FlucSpecTrans}'s left side, and encompasses the typical set, as a
consequence of our design. Rather nearby the Engine regime, though, is a narrow
one of no functioning at all---a Dud. In fact, though the ratchet was designed
as an Engine, we see that over most of the fluctuations, with the given
parameter setting the ratchet operates as an Eraser.

Finally, App. \ref{app:MaxWorkIndepInputBias} shows that the maximum work, over
all fluctuation subsets---all $\beta$ or all allowed $U$s---is independent of
the input process bias. This is perhaps puzzling as bias clearly controls the
ratchet's thermodynamic behavior. Thus, assuming an IID input, the maximum work
is a property of the ratchet itself and not the input, playing a role rather
analogous to how Shannon's channel capacity is a channel property.

\subsection{Probable Functional Fluctuations}
\label{sec:ProbFuncFluc}

How probable are fluctuations in thermodynamic function? The answer, at first
sight, is not entirely obvious, given that we are asking a question about
deviations from the typical set and so are asking about the likelihood of a
property of rare realizations. Indeed, statistical variations in this or that
property might not be practically observable at all. We now show that the
functional fluctuations are, in fact, quite observable even at relatively
long word lengths, such as $\ell = 100$.

To answer this we first need to address how likely we are to observe a
fluctuation. The large-deviation rate function $I(U)$ provides the answer as it
gives the probability of the subset of sequences with the same energy $U$. The
Gartner-Ellis theorem \cite{Bowe75a,Buck90a,Touc09a} says that the probability
of a sequence occurring in a fluctuation set with energy density $U$ is determined by:
\begin{align*}
I(U) = \lim_{\ell \to \infty}
  -\frac{\log_2 \Pr (U^{\ell})}{\ell}
  ~.
\end{align*}
That is, the subset probability scales exponentially: $\Pr (U^{\ell}=U) =
\exp(-I(U)\ell) + \mathscr{O}(\ell)$, where $\mathscr{O}(\cdot)$ decays faster than any exponential.

Importantly, we can directly determine the large-deviation rate function $I(U)$
as it is directly related to the fluctuation spectrum $S(U)$ just derived in
Eqs. (\ref{eq:EnergyDensity}) and (\ref{eq:EntropyDensity}) \cite{Agha15a}:
\begin{align}
I(U) = U - S(U)
  ~.
\label{eq:RateFunction}
\end{align}

To understand this a bit more, let's compare to $S(U(\beta))$. For large
$\ell$, as noted above, $\beta=1$ indicates the typical set, its sequences'
probabilities decay at the entropy rate $\hmu$ and the probability of observing
a realization in the typical set converges to $1$. Thus, $I(U) = 0$ there and
$S(U) = U$. For other fluctuation sets with energy density $U$, we expect the
probability of the fluctuation subset at $U$ to vanish with increasing
length $\ell$ and $I(U)$ indicates exactly how fast this decay is.

\begin{figure}
\centering
\includegraphics[width=\columnwidth]{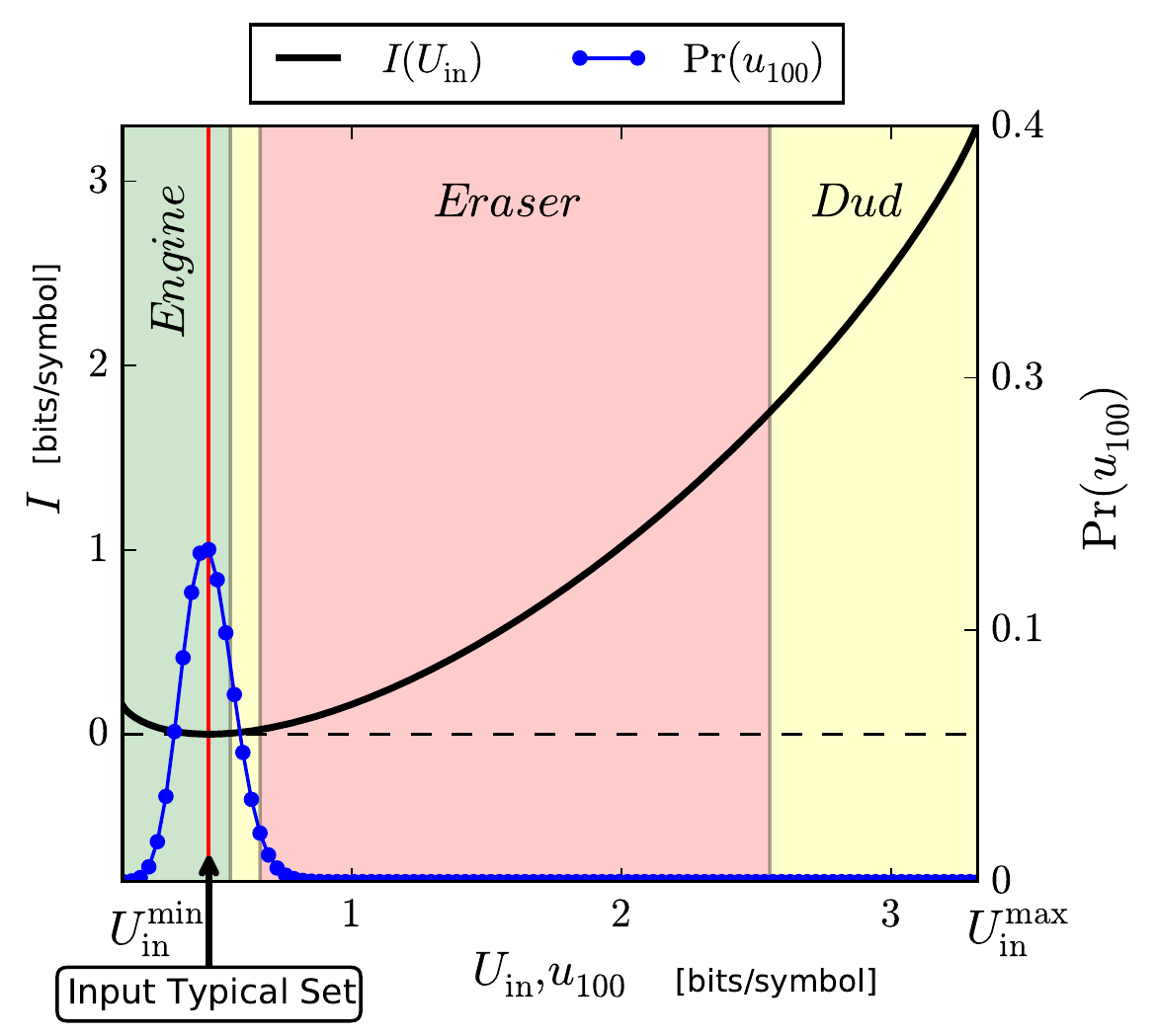}
\caption{Decay rate of probability of fluctuations in thermodynamic
  functioning: Large-deviation rate function $I(U_\text{in})$ (solid black
  line) and the probability $\Pr(u_{100})$ of fluctuation subsets for length
  $\ell = 100$ input realizations (dotted blue line).
  }
\label{fig:thermofluciu}
\end{figure}

So, now we can ask how likely the ratchet is to fluctuate between its possible
thermodynamic modalities. This is determined from the large deviation rate
function $I(U)$ of Eq. (\ref{eq:RateFunction}), which
Fig.~\ref{fig:thermofluciu} plots as a function of $U_\text{in}$. As the figure
shows, when realizations from the typical set are fed in, the ratchet functions
as an Engine. Also, this subset happens with zero large deviation rate. At the
limit of infinite length the probability of the typical set goes to one and the
probability of fluctuation subsets vanishes. The ratchet operates as an engine
over long times with probability one. In reality, though, we only work with
finite length sequences. And so, the operant question here is, are these
functional fluctuations observable at finite lengths? As we alluded to much
earlier, short sequences enhance their observation.

Consider the input process in Fig.~\ref{fig:Machines}(a) and assume the input's
realization length is $\ell = 100$. For this case we have $2^{100}$ distinct
input sequences that are partitioned into $101$ fluctuation subsets with
different energy densities---subsets of sequences with $n$ $0$s and $100-n$
$1$s for $n = 0, 1, \ldots, 100$. Let's calculate the probability of each of
these fluctuations subsets occurring. The probability of each versus its energy
is shown in Fig.~\ref{fig:thermofluciu} as the blue dotted line. To distinguish
it from the energy density of fluctuation subsets at infinite length we label
the energy density of each of these sets with $u_{100}$, the index $100$
reminds us that we are examining input sequences of length $\ell = 100$. There
are $101$ blue points on the figure, each representing one of the fluctuation
subsets. From $101$ fluctuation subsets, if we fed $13$ of them (the first 13
blue points in the left of the figure) to the ratchet, the ratchet functions as
an Engine. This means for the other $87$ fluctuation subsets the ratchet
functions as a Dud or Eraser. By calculating the probabilities we see by
feeding an input sequence with length $100$, with approximately $80\%$
probability the ratchet functions as an Engine, with approximately $17.8\%$
probability it functions as a Dud, and with $2.2\%$ probability functions as an
Eraser.

\section{Conclusion}

We synthesized statistical fluctuations---as entailed in Shannon's Asymptotic
Equipartition Property \cite{Cove06a} and large deviation theory
\cite{Bowe75a,Buck90a,Touc09a}---and functional thermodynamics---as determined
using the new informational Second Law \cite{Boyd15a}---to predict spontaneous
variations in thermodynamic functioning. In short, there is simultaneous,
inherently parallel, thermodynamic processing that is functionally distinct and
possibly in competition. This strongly suggests that, even when in a
nonequilibrium steady state, a single nanoscale device or biomolecule can be
both an engine and an eraser. And, we showed that these functional fluctuations
need not be rare. The conclusion is that functional fluctuations should be
readily observable and the prediction experimentally testable.

A main point motivating this effort was to call into question the widespread
habit of ascribing a single functionality to a given system and, once that veil
has lifted, to appreciate the broad consequences. To drive them home, since
biomolecular systems are rather like the information ratchet here, they should
exhibit, measurably different thermodynamic functions as they behave. If this
prediction holds, then the biological world is vastly richer than we thought
and it will demand of us a greatly refined vocabulary and greatly improved
theoretical and experimental tools to adequately probe and analyze this new
modality of parallel functioning.

That said, thoroughness forces us to return to our earlier caveat (Sec.
\ref{sec:FuncFluc}) concerning not conflating various ``temperatures". If we
give the input information reservoir and the output information reservoir
physical implementations, then the fluctuation indices $U_\text{in}$ and
$U_\text{out}$ take on thermal physical meaning and so can be related to the
ratchet's thermodynamic temperature $T$. Doing so, however, would take us too
far afield here, but it will be necessary for a complete understanding.

Equally important, the theoretical scaffolding used above in the service of
illustrating parallel thermodynamic functioning invokes a number of
simplifications. Perhaps the main one is the use of time-asymptotic quantities,
such as the Shannon entropy rate and large deviation rate function. A proper
analysis requires carefully working in the finite-time, finite-length sequence
regime---the very regime that enhances statistical fluctuations. This task is
markedly more challenging and will be attempted elsewhere. However, the central
goal has been to explicate the main ideas and these are robust to the
simplifications employed. Moreover, we drove the ratchet with atypical input
sequences, assuming that the ratchet responded typically. However, in addition,
we could have explored the ratchet's atypical behavior in response to input
typical sequences. Or both: analyze the atypical transduction of atypical
inputs.

Similarly looking forward, there are sister challenges. First, note that
technically speaking we introduced a fluctuation theory for memoryful stochastic
transducers, but by way of the example of Ref. \cite{Boyd15a}'s information
ratchet. A thoroughgoing development must be carried out in much more
generality using the tools of Refs. \cite{Barn13a}, \cite{Agha15a}, and
\cite{Boyd16c}, if we are to fully understand the functionality of
thermodynamic processes that transform inputs to outputs, environmental
stimulus to environmental action.

Second, the role of Jarzynski-Crooks theory for fluctuations in thermodynamic
observables needs to be made explicit and directly related to statistical
fluctuations, in the sense emphasized here. One reason is that their theory
bears directly on controlling thermodynamic systems and the resulting
macroscopic fluctuations. To draw the parallel more closely, we could drive the
ratchet parameters $p$ and $q$ and input bias $b$ between different functional
regimes and monitor the entropy production fluctuations to test how the theory
fares for memoryful processes. In any case, efficacy in control will also be
modulated by statistical fluctuations.

Not surprisingly, there is much to do. Let's turn to a larger motivation and
perhaps larger consequences to motivate future efforts.

As just noted, fluctuations are key to nanoscale physics and molecular
biology.  We showed that fluctuations are deeply implicated both in identifying
thermodynamic function and in the very operation of small-scale systems. In
fact, fluctuations are critical to life---its proper and robust functioning.
The perspective arising from parallel thermodynamic function is that, rather
than fluctuations standing in contradiction to life processes, potentially
corrupting them, there may be a positive role for fluctuations and parallel
thermodynamic functioning. Once that is acknowledged it is a short step to
realize that biological evolution \cite{Crut99d} may have already harnessed
them to good thermodynamic effect. Manifestations are clearly worth looking for.

It now seems highly likely that fluctuations engender more than mere health and
homeostasis. It is a commonplace that biological evolution is nothing, if not
opportunistic. If so, then it would evolve cellular biological thermodynamic
processes that actively leverage fluctuations. Mirroring Maxwell's Demon's need
for fluctuations to operate, biological evolution itself advances only when
there are fluctuations. For example, biomolecular mutation processes engender a
distribution of phenotypes and fitnesses; fodder for driving selection and so
evolutionary innovation. This, then, is \emph{Darwin's Demon}---a mechanism
that ratchets in favorable fluctuations for positive thermodynamic and then
positive survival benefit. The generality of results and methods here give new
insight into thermodynamic functioning in the presence of fluctuations that
should apply at many different scales of life, including its emergence and
evolution.



\section*{Acknowledgments}
\label{sec:acknowledgments}

We thank Alec Boyd, John Mahoney, Dibyendu Mandal, Sarah Marzen, and Paul
Riechers for helpful discussions. JPC thanks the Santa Fe Institute for its
hospitality during visits as an External Faculty member. This material is based
upon work supported by, or in part by, the John Templeton Foundation and U. S.
Army Research Laboratory and the U. S. Army Research Office under contracts
W911NF-13-1-0390 and W911NF-13-1-0340.

\appendix

\section{Process Fluctuations from the Twisted \EM}
\label{app:FlucFromEM}

A process' fluctuation spectrum is calculated from its twisted \eM. Introducing
the latter requires briefly recalling several important and relevant concepts.
Though the following closely tracks Refs. \cite{Youn93a,Agha15a}, the synopsis
here is relatively self contained as far as basic calculations are concerned.

The \emph{Shannon block entropy} is a linear average of the sequence
self-informations $-\log_2 \Prob(w)$. The closely related \emph{Renyi block
entropy} is the most general entropy that is both additive over independent
distributions (extensive) and a geometric average \cite{Camp65a}:
\begin{align}
\H_\beta(\ell) & = \H_\beta[X_{0:\ell}] \nonumber \\
 & = \frac {1}{1-\beta} \log_2 \sum_{\{w \in \Abet^\ell\}} (\Prob(w))^\beta
  ~,
\label{eq:RBlockEntropy}
\end{align}
where $\beta$ is an arbitrary real number that allows us to ``focus'' on
sequence subsets parametrized by probability---or, equivalently, by energy $U_w
= - \log \Pr(w)$. We see that $\beta$ is analogous to inverse temperature and
we can interpret the sum $ \sum_{\{w \in \Abet^\ell\}} (\Prob(w))^\beta$ as the
\emph{partition function}:
\begin{align}
\mathcal{Z}_{\ell}(\beta) = \sum_{\{w \in \Abet^\ell\}} e^{- \beta (- \ln \Prob(w))}
  ~.
\label{eq:Partit}
\end{align}

Now, we are ready to define a process' twisted \eM, which is determined from
the process' \eM. It is the analog of the \emph{escort} or \emph{twisted}
distributions of large deviation theory \cite{Bowe75a,Buck90a,Touc09a}, but
adapted to our setting of structured processes.

\begin{Def}\label{newproc}
A process $\Process$'s \emph{twisted \eM} is the parametrized family of \eMs\
$M_\beta = \big\{ \CausalStateSet, \{ \textbf{S}_\beta^{(\ms)}: \ms \in
\MeasAlphabet \}, \bra{\widehat{\eta}_0} \big\} $, where the components are the
same as $\Process$'s \eM\ $M(\Process)$, except that there is an inverse
temperature parameter $\beta$ and a new, parametrized transition dynamic:
\begin{align}
\big( \textbf{S}^{(\ms)}_\beta \big)_{ij}
  = \frac {\big(\textbf{T}^{(\ms)}_\beta \big)_{ij} (\MaxRvecBeta)_j }
  {\MaxEigBeta (\MaxRvecBeta)_i }
  ~.
\label{eq:Sxij}
\end{align}
$\Process$'s \eM\ transition matrices $T_{ij}^{(x)} = \Pr(\cs_j,x \vert \cs_i)$ are transformed to:
\begin{align*}
\big( \textbf{T}^{(\ms)}_\beta \big)_{ij}
  & = e^{\beta \ln{\Prob(\cs_j,\ms \vert \cs_i)}} \\
  & =  \big( \Prob(\cs_j,\ms \vert \cs_i) \big)^\beta
  ~.
\end{align*}
We calculate its eigenvectors and eigenvalue as follows. Form $M_\beta$'s
internal causal-state transition matrix:
\begin{align*}
\textbf{T}_\beta = \sum_{\ms \in \Abet}  T^{(\ms)}_{\beta}
  ~.
\end{align*}
Then $\MaxLvecBeta$ ($\MaxRvecBeta$) is the left (right) eigenvector of
$\textbf{T}_\beta$, associated with $\MaxEigBeta$:
\begin{align*}
\MaxLvecBeta \textbf{T}_\beta & = \MaxEigBeta \MaxLvecBeta \\
\textbf{T}_\beta \MaxRvecBeta & = \MaxEigBeta \MaxRvecBeta
  ~,
\end{align*}
where $\MaxEigBeta$ is $\textbf{T}_{\beta}$'s maximum eigenvalue. We chose the
eigenvectors such that:
\begin{align}
\MaxLvecBeta \cdot \MaxRvecBeta = 1
  ~.
\label{eq:Normal}
\end{align}
The new initial state probability distribution $\bra{\widehat{\eta}_0}$ is the
normalized left eigenvector of $\sum_{\{\ms\}} {\textbf S}^{(\ms)}_\beta$.
\end{Def}

Let $\Pr(w)$ denote the probability that the original \eM\ $M$ generates the
length-$\ell$ sequence $w$. Then the probability of the same sequence being
generated by $M_\beta$ is \cite{Youn93a,Agha15a}:
\begin{align}
\mathcal{Q}_{\beta}(w)=\frac{(\Prob(w))^\beta}{\mathcal{Z}(\beta)}
  ~.
\end{align}
It can be shown \cite{Youn93a,Agha15a} that generating the \emph{twisted
distribution} $\mathcal{Q}_{\beta}(\cdot)$ is equivalent to generating a new
process whose typical set is the fluctuation subset at $U=U(\beta)$ in the
original process, where:
\begin{align}
U(\beta) = \frac{1}{\beta} ( \MetricER(M_\beta) -  \log \MaxEigBeta )~.
\label{eq:SUHRELATION}
\end{align}
This relates the energy and entropy densities, since the latter monitors the
set's growth rate; that is, $S(U) = \MetricER(M_\beta)$. Thus, by varying
$\beta$ we choose which fluctuation subset $U(\beta)$ to focus in on.
Critically, though, we have a new \eM\ $M_\beta$ for which that subset is
typical and so generated with high probability.

The fluctuation subset parameter $\beta$ is $S(U)$'s slope, illustrating that
$S(U)$ is well behaved. Proof of convexity is found in the references cited
above.

{\The $\beta = \text{d} S / \text{d} U$.}
{\ProThe
\begin{align}\label{dl}
\frac{{\rm d}\MaxEigBeta}{{\rm d}\beta}&=\sum_{i,j}
(\MaxLvecBeta)_j
\frac{{\rm d}({\textbf T}_{\beta})_{ij}}{{\rm d}\beta}
(\MaxRvecBeta)_j\nonumber\\
&=\frac{1}{\beta}\sum_{i,j}
(\MaxLvecBeta)_j
({\textbf T}_{\beta})_{ij}
(\MaxRvecBeta)_j\log ({\textbf T}_{\beta})_{ij}.
\end{align}
From Eqs. (\ref{eq:Sxij}) and  (\ref{eq:Normal}) one sees that:
\begin{equation*}
(\MaxLvecBeta)_j
({\textbf T}_{\beta})_{ij}
(\MaxRvecBeta)_j=\MaxEigBeta ({\textbf P_{\beta}})_i({\textbf S_{\beta}})_{ij}
  ~,
\end{equation*}
which defines $\textbf{P}_\beta$.
Using this in Eq. (\ref{dl}) gives:
\begin{align}
\label{dl2}
\frac{1}{\MaxEigBeta} \frac{{\rm d}\MaxEigBeta}{{\rm d}\beta}
  & = \frac{1}{\beta}\sum_{i,j}
 ({\textbf P_{\beta}})_i({\textbf S_{\beta}})_{ij}
 \log({\textbf T}_{\beta})_{ij}\nonumber\\
 & = \frac{1}{\beta}\sum_{i,j}
  ({\textbf P_{\beta}})_i({\textbf S_{\beta}})_{ij}
   \left[\log ({\textbf S_{\beta}})_{ij}+\log\MaxEigBeta\right.
    \nonumber\\
   & \quad\quad \left.+\log(\MaxRvecBeta)_i-(\MaxRvecBeta)_j  \right]
   \nonumber\\
  &= -\frac{S(U(\beta))}{\beta}+\frac{\log{\MaxEigBeta}}{\beta}
  \nonumber \\
  & \quad +\frac{1}{\beta}\sum_{i,j}
    ({\textbf P_{\beta}})_i({\textbf S_{\beta}})_{ij}\left[\log(\MaxRvecBeta)_i-(\MaxRvecBeta)_j  \right]
  .
\end{align}
To obtain the first term above, the definition of entropy is
used and for the second term one makes use of:
\begin{align*}
\nonumber\sum_{i,j} ({\textbf P_{\beta}})_i ({\textbf S_{\beta}})_{ij}
\nonumber  & = \sum_{j}({\textbf P_{\beta}})_j \\
\nonumber  & = \sum_{j} (\MaxRvecBeta)_j
  (\MaxLvecBeta)_j \\
\nonumber  & = {\textbf l_{\beta}}\cdot{\textbf r_{\beta}} \\
  & = 1
  ~.
\end{align*}
Now, using:
\begin{align*}
\sum_{i} ({\textbf P_{\beta}})_i ({\textbf S_{\beta}})_{ij}=({\textbf P_{\beta}})_j
  = (\MaxRvecBeta)_j (\MaxLvecBeta)_j  
  ~,
\end{align*}
the third and the fourth terms in Eq. (\ref{dl2}) simply cancel, and
one arrives at:
\begin{equation}\label{d2}
\frac{{\rm d} }{{\rm d}\beta}\,(\log \MaxEigBeta)= -U(\beta)
  ~.
\end{equation}
Then one may take $S(\cdot)$ as a function of $\beta$. Multiplying both sides of
Eq.~(\ref{eq:SUHRELATION}) by $\beta$ and differentiating both sides with respect to $\beta$,
one obtains:
\begin{align*}
U + \beta\frac{{\rm d}U}{{\rm d}\beta}= \frac{{\rm d}S}{{\rm d}\beta}- \frac{{\rm d} }{{\rm d}\beta}\,(\log \MaxEigBeta),
\end{align*}
Using Eq.~(\ref{d2}), one finds:
\begin{align}\label{SMAX}
\nonumber\beta & = \frac{{{\rm d}S}/{{\rm d}\beta}}{{{\rm d}U}/{{\rm d}\beta}} \\
  & =\frac{{\rm d}S}{{\rm d}U}
  ~.
\end{align}
Thus, $\beta$ indeed plays the same role here as the inverse temperature in
statistical physics. One consequence of Eq.~(\ref{SMAX}) is that $S(\cdot)$ is
a well behaved function. From Def. \ref{newproc} the typical set is found at
$\beta=1$. And, this means that at the typical set we have $\mathrm{d} S /
\mathrm{d} U=1$.
}

\section{Biased Coin Fluctuation Spectrum}
\label{app:BCFlucSpec}

First, recall Eq. (\ref{eq:SUHRELATION}):
\begin{align*}
S(U(\beta)) = \beta U(\beta) + \log \lambda_\beta
  ~.
\end{align*}
Calculating the maximal eigenvalue $\widehat{\lambda}_\beta$, we find:
\begin{align*}
\log \lambda_\beta = \log_2(b^\beta + (1-b)^\beta)
  ~.
\end{align*}
Second, for the entropy density recall that:
\begin{align*}
S(U(\beta)) = \hmu(M_\beta)
  ~.
\end{align*}
Substituting $M_\beta$ bias $\widehat{b} = b^\beta / (b^\beta +
(1-b)^\beta$ into Eq. (\ref{eq:InputEntropyRate}), we find:
\begin{align}
S(U(\beta)) =
  - & \left( \frac{b^\beta}{b^\beta + (1-b)^\beta} \log_2 \frac{b^\beta}{b^\beta + (1-b)^\beta} \right.
  \nonumber \\
  & \left. + \frac{(1-b)^\beta}{b^\beta + (1-b)^\beta} \log_2 \frac{(1-b)^\beta}{b^\beta + (1-b)^\beta} \right)
  ~.
\label{eq:EntropyBeta}
\end{align}
It is straightforward, now, to calculate $U$ from these:
\begin{align}
U(\beta) & =
  \frac{-b^\beta}{b^\beta + (1-b)^\beta} \log_2 (b) \nonumber \\
  & \quad\quad + \frac{-(1-b)^\beta}{b^\beta + (1-b)^\beta} \log_2 (1-b)
  ~.
\label{eq:EnergyBeta}
\end{align}
Plotting Eq. (\ref{eq:EntropyBeta})
against Eq. (\ref{eq:EnergyBeta}) gives the biased coin fluctuation
spectrum shown in Fig. \ref{fig:FlucSpecTrans}.

\section{Typical Set for a Biased Coin}
\label{TSBC}

What is $A_\epsilon^{(\ell)}$ for a biased coin with bias $b$? The typical set
is defined by;
\begin{align*}
  A_\epsilon^{(\ell)} = \lbrace
  w \in \Abet^\ell :
  2^{- \ell (\hmu+\epsilon)} \leq \Pr(w) \leq 2^{- \ell ( \hmu - \epsilon)}  \rbrace~.
\end{align*}
The probability of a biased coin generating a particular sequence $w$ with $k$
heads is $b^k(1-b)^{(\ell-k)}$. And so, for $w$ to be in the typical set we
must have:
\begin{align*}
\ell b - \frac{\ell \epsilon}{\log{\frac{b}{1-b}}} \leq k \leq \ell b +
\frac{\ell \epsilon}{\log{\frac{b}{1-b}}}~.
\end{align*}
Since $k$ is an integer:
\begin{align*}
\ceil*{\ell b - \frac{\ell \epsilon}{\log{\frac{b}{1-b}}}} \leq k \leq
\floor*{\ell b + \frac{\ell \epsilon}{\log{\frac{b}{1-b}}}}~.
\end{align*}

For example, in the case where $\ell=1000$, $b=0.6$, and $\epsilon = 0.01$, we
have:
\begin{align*}
582 \leq k \leq 617~.
\end{align*}
This means that those length $\ell = 1000$ sequences with $582$ to
$617$ Heads are in the typical set.

\section{Maximum Work is Independent of Input Process}
\label{app:MaxWorkIndepInputBias}

\newcommand{\Work}{\langle W \rangle}
\newcommand{\Wmax}{\widehat{W}}

The maximum work $\Wmax$ done by the information ratchet over all fluctuation
subsets (parametrized by $\beta$, say) is independent of the given IID binary
input process. Direct calculation gives:
\begin{align*}
\Wmax & = \max_{\beta} \Work \\
      & = \begin{cases}
			\Wmax^- \equiv \kB T \big(-q \log (p/q) - q \log (1-q) \big) & c < 0 \\
			\Wmax^+ \equiv \kB T \big( p \log (p/q) - p \log (1-p) \big) & c \geq 0 \\
          \end{cases}
  ~.
\end{align*}
where $c = (p+q) \log (q/p) + p \log (1-p) - q \log (1-q)$.

\section{Work is Linear in Energy Density}
\label{app:WorkLinearInU}

Recall the energy density $U$ parametrizes the fluctuation subsets. Here, we
show that the work $\Work$ is linear across the $U$-fluctuation classes:
\begin{align*}
\Work (U) = \frac{\kB T}{2} \left( c U + U_0 \right)
   ~,
\end{align*}
where:
\begin{align*}
U_0 & = \tfrac{1}{2} \Wmax^- + c \times \frac{\log(1-b)}{\log (1-b) - \log b}
  ~.
\end{align*}

To see this, we first calculate the work $W(\beta)$ from Eq. (\ref{eq:Work}):
\begin{align*}
W(\beta) = \frac{\kB T}{2}
  \left( -q \log (q/p) + q \log (1-q) + \frac{c b^\beta}{b^\beta + (1-b)^\beta}
  \right)
  .
\end{align*}
Now, for $W$ in terms of $U$ we find:
\begin{align*}
W(U) = \frac{\kB T}{2}
  & \left( -q \log \left(\frac{q}{p}\right) + q \log (1-q) \big) \right. \\
  & \left. + c \frac{U + \log(1-b)}{\log(1-b) - \log(b)} \right)
  ~,
\end{align*}
which is the linear form claimed.

\bibliography{chaos}

\end{document}

%% file: cmechabbrev.tex
\newcommand{\eM}     {\mbox{$\epsilon$-machine}}
\newcommand{\eMs}    {\mbox{$\epsilon$-machines}}
\newcommand{\EM}     {\mbox{$\epsilon$-Machine}}
\newcommand{\EMs}    {\mbox{$\epsilon$-Machines}}
\newcommand{\eT}     {\mbox{$\epsilon$-transducer}}

\newcommand{\ETs}    {\mbox{$\epsilon$-Transducers}}

\newcommand{\Process}{\mathcal{P}}

\newcommand{\MeasAlphabet}  {\mathcal{A}}
\newcommand{\MeasSymbol}   { {X} }
\newcommand{\meassymbol}   { {x} }

\newcommand{\CausalState}   { \mathcal{S} }

\newcommand{\causalstate}   { \sigma }
\newcommand{\CausalStateSet}    { \boldsymbol{\CausalState} }

\newcommand{\Prob}      {\Pr} 

\newcommand{\hmu}       {h_\mu}

\newcommand{\ProcessAlphabet}   {\MeasAlphabet}

\newcommand{\forward}{+}
\newcommand{\reverse}{-}
\newcommand{\forwardreverse}{\pm} 

\newcommand{\FutureCausalState} { {\CausalState}^{\forward} }

\newcommand{\PastCausalState}   { {\CausalState}^{\reverse} }

\newcommand{\lastindex}[2]{
  \edef\tempa{0}
  \edef\tempb{#2}
  \ifx\tempa\tempb
    \edef\tempc{#1}
  \else
    \edef\tempa{0}
    \edef\tempb{#1}
    \ifx\tempa\tempb
      \edef\tempc{#2}
    \else
      \edef\tempc{#1+#2}
    \fi
  \fi
  \tempc
}

\newcommand{\CSjoint}[1][,]{
   \edef\tempa{:}
   \edef\tempb{#1}
   \ifx\tempa\tempb
      \ensuremath{\FutureCausalState\!#1\PastCausalState}
   \else
      \ensuremath{\FutureCausalState#1\PastCausalState}
   \fi
}

\newif\ifpm
\edef\tempa{\forwardreverse}
\edef\tempb{\pm}
\ifx\tempa\tempb
   \pmfalse
\else
   \pmtrue
\fi

%% file: naface.bbl
\begin{thebibliography}{10}

\bibitem{Szil29a}
L.~Szilard.
\newblock On the decrease of entropy in a thermodynamic system by the
  intervention of intelligent beings.
\newblock {\em Z. Phys.}, 53:840--856, 1929.

\bibitem{Maxw88a}
J.~C. Maxwell.
\newblock {\em Theory of Heat}.
\newblock Longmans, Green and Co., London, United Kingdom, ninth edition, 1888.

\bibitem{Boyd14b}
A.~B. Boyd and J.~P. Crutchfield.
\newblock Demon dynamics: {Deterministic} chaos, the {Szilard} map, and the
  intelligence of thermodynamic systems.
\newblock {\em Phys. Rev. Lett.}, 116:190601, 2016.

\bibitem{Bril62a}
L.~Brillouin.
\newblock {\em Science and Information Theory}.
\newblock Academic Press, New York, second edition, 1962.

\bibitem{Toya10a}
S.~Toyabe, T.~Sagawa, M.~Ueda, E.~Muneyuki, and M.~Sano.
\newblock Experimental demonstration of information-to-energy conversion and
  validation of the generalized {Jarzynski} equality.
\newblock {\em Nat. Physics}, 6:988--992, 2010.

\bibitem{Lamb11a}
B.~Lambson, D.~Carlton, and J.~Bokor.
\newblock Exploring the thermodynamic limits of computation in integrated
  systems: Magnetic memory, nanomagnetic logic, and the {Landauer} limit.
\newblock {\em Phys. Rev. Lett.}, 107:010604, 2011.

\bibitem{Beru2012}
A.~Berut, A.~Arakelyan, A.~Petrosyan, S.~Ciliberto, R.~Dillenschneider, and
  E.~Lutz.
\newblock Experimental verification of {Landauer's} principle linking
  information and thermodynamics.
\newblock {\em Nature}, 483:187, 2012.

\bibitem{Jun14a}
Y.~Jun, M.~Gavrilov, and J.~Bechhoefer.
\newblock High-precision test of {Landauer's} principle.
\newblock {\em Phys. Rev. Lett.}, 113:190601, 2014.

\bibitem{Mada14a}
M.~Madami, M.~d'YAquino, G.~Gubbiotti, S.~Tacchi, C.~Serpico, and G.~Carlotti.
\newblock Micromagnetic study of minimum-energy dissipation during {Landauer}
  erasure of either isolated or coupled nanomagnetic switches.
\newblock {\em Phys. Rev. B}, 90:104405, 2014.

\bibitem{Peko15a}
J.~P. Pekola.
\newblock Towards quantum thermodynamics in electronic circuits.
\newblock {\em Nat. Physics}, 11:118--123, 2015.

\bibitem{Kosk15a}
J.~V. Koski, A.~Kutvonen, I.~M. Khaymovich, T.~Ala-Nissila, and J.~P. Pekola.
\newblock On-chip {Maxwell's} demon as an information-powered refrigerator.
\newblock {\em Phys. Rev. Lett.}, 115:260602, 2015.

\bibitem{Hong16a}
J.~Hong, B.~Lambson, S.~Dhuey, and J.~Bokor.
\newblock Experimental test of {Landauer's} principle in single-bit operations
  on nanomagnetic memory bits.
\newblock {\em Sci. Adv.}, 2:e1501492, 2016.

\bibitem{Evan93a}
D.~J. Evans, E.~G.~D. Cohen, and G.~P. Morriss.
\newblock Probability of second law violations in shearing steady flows.
\newblock {\em Phys. Rev. Lett.}, 71:2401--2404, 1993.

\bibitem{Evan1994}
D.~J. Evans and D.~J. Searles.
\newblock Equilibrium microstates which generate second law violating steady
  states.
\newblock {\em Phys. Rev. E}, 50:1645, 1994.

\bibitem{Gall95a}
G.~Gallavotti and E.~G.~D. Cohen.
\newblock Dynamical ensembles in nonequilibrium statistical mechanics.
\newblock {\em Phys. Rev. Lett.}, 74:2694--2697, 1995.

\bibitem{Kurc1998}
J.~Kurchan.
\newblock Fluctuation theorem for stochastic dynamics.
\newblock {\em J. Phys. A: Math. Gen.}, 31:3719, 1998.

\bibitem{Croo98a}
G.~E. Crooks.
\newblock Nonequilibrium measurements of free energy differences for
  microscopically reversible {Markovian} systems.
\newblock {\em J. Stat. Phys.}, 90(5/6):1481--1487, 1998.

\bibitem{Lebo1999}
J.~L. Lebowitz and H.~Spohn.
\newblock A {Gallavotti-Cohen}-type symmetry in the large deviation functional
  for stochastic dynamics.
\newblock {\em J. Stat. Phys.}, 95:333, 1999.

\bibitem{Coll2005}
D.~Collin, F.~Ritort, C.~Jarzynski, S.~B. Smith, I.~Tinoco Jr., and
  C.~Bustamante.
\newblock Verification of the {Crooks} fluctuation theorem and recovery of
  {RNA} folding free energies.
\newblock {\em Nature}, 437:231, 2005.

\bibitem{Croo99a}
G.~E. Crooks.
\newblock Entropy production fluctuation theorem and the nonequilibrium work
  relation for free energy differences.
\newblock {\em Phys. Rev. E}, 60:2721, 1999.

\bibitem{Liph02a}
J.~Liphardt, S.~Dumont, S.~B. Smith, I.~Tinoco, and C.~Bustamante.
\newblock Equilibrium information from nonequilibrium measurements in an
  experimental test of {Jarzynski's} equality.
\newblock {\em Science}, 296:1832, 2002.

\bibitem{Coll05a}
D.~Collin, F.~Ritort, C.~Jarzynski, S.~B. Smith, I.~Tinoco, and C.~Bustamante.
\newblock Verification of the {Crooks} fluctuation theorem and recovery of
  {RNA} folding free energies.
\newblock {\em Nature}, 437:231, 2005.

\bibitem{Alem12a}
A.~Alemany, A.~Mossa, I.~Junier, and F.~Ritort.
\newblock Experimental free-energy measurements of kinetic molecular states
  using fluctuation theorems.
\newblock {\em Nat. Physics}, 8:688--694, 2012.

\bibitem{Boyd15a}
A.~B. Boyd, D.~Mandal, and J.~P. Crutchfield.
\newblock Identifying functional thermodynamics in autonomous {Maxwellian}
  ratchets.
\newblock {\em New J. Physics}, 18:023049, 2016.

\bibitem{Mand012a}
D.~Mandal and C.~Jarzynski.
\newblock Work and information processing in a solvable model of {Maxwell's}
  demon.
\newblock {\em Proc. Natl. Acad. Sci. USA}, 109(29):11641--11645, 2012.

\bibitem{Varn15a}
D.~P. Varn and J.~P. Crutchfield.
\newblock What did {Erwin} mean? {The} physics of information from the
  materials genomics of aperiodic crystals and water to molecular information
  catalysts and life.
\newblock {\em Phil. Trans. Roy. Soc. A}, 374:20150067, 2016.
\newblock In Theme Issue on ``DNA as information: {At} the crossroads between
  biology, mathematics, physics and chemistry''.

\bibitem{Mand2013}
D.~Mandal, H.~T. Quan, and C.~Jarzynski.
\newblock Maxwell's refrigerator: {An} exactly solvable model.
\newblock {\em Phys. Rev. Lett.}, 111:030602, 2013.

\bibitem{Stra2013}
P.~Strasberg, G.~Schaller, T.~Brandes, and M.~Esposito.
\newblock Thermodynamics of a physical model implementing a {Maxwell} demon.
\newblock {\em Phys. Rev. Lett.}, 110:040601, 2013.

\bibitem{Bara2013}
A.~C. Barato and U.~Seifert.
\newblock An autonomous and reversible {Maxwell's} demon.
\newblock {\em Europhys. Lett.}, 101:60001, 2013.

\bibitem{Hopp2014}
J.~Hoppenau and A.~Engel.
\newblock On the energetics of information exchange.
\newblock {\em Europhys. Lett.}, 105:50002, 2014.

\bibitem{Lu14a}
Z.~Lu, D.~Mandal, and C.~Jarzynski.
\newblock Engineering {Maxwell's} demon.
\newblock {\em Physics Today}, 67(8):60--61, January 2014.

\bibitem{Um2015}
J.~Um, H.~Hinrichsen, C.~Kwon, and H.~Park.
\newblock Total cost of operating an information engine.
\newblock {\em arXiv:1501.03733 [cond-mat.stat-mech]}, 2015.

\bibitem{Crut88a}
J.~P. Crutchfield and K.~Young.
\newblock Inferring statistical complexity.
\newblock {\em Phys. Rev. Let.}, 63:105--108, 1989.

\bibitem{Land61a}
R.~Landauer.
\newblock Irreversibility and heat generation in the computing process.
\newblock {\em IBM J. Res. Develop.}, 5(3):183--191, 1961.

\bibitem{Benn82}
C.~H. Bennett.
\newblock Thermodynamics of computation - a review.
\newblock {\em Intl. J. Theo. Phys.}, 21:905, 1982.

\bibitem{Crut12a}
J.~P. Crutchfield.
\newblock Between order and chaos.
\newblock {\em Nature Physics}, 8(January):17--24, 2012.

\bibitem{Barn13a}
N.~Barnett and J.~P. Crutchfield.
\newblock Computational mechanics of input-output processes: {Structured}
  transformations and the $\epsilon$-transducer.
\newblock {\em J. Stat. Phys.}, 161(2):404--451, 2015.

\bibitem{Cove06a}
T.~M. Cover and J.~A. Thomas.
\newblock {\em Elements of Information Theory}.
\newblock Wiley-Interscience, New York, second edition, 2006.

\bibitem{Note1}
This generalizes \cite {Youn93a,Agha15a} the scaling for memoryless processes
  (independent, identically distributed) presented in, for example, Ref. \cite
  [Ch. 3]{Cove06a}.

\bibitem{Shan48a}
C.~E. Shannon.
\newblock A mathematical theory of communication.
\newblock {\em Bell Sys. Tech. J.}, 27:379--423, 623--656, 1948.

\bibitem{McMi53a}
B.~McMillan.
\newblock The basic theorems of information theory.
\newblock {\em Ann. Math. Stat.}, 24:196--219, 1953.

\bibitem{Brei57}
L.~Breiman.
\newblock The individual ergodic theorem of information theory.
\newblock {\em Ann. Math. Stat.}, 28(3):809--811 

\bibitem{Youn93a}
K.~Young and J.~P. Crutchfield.
\newblock Fluctuation spectroscopy.
\newblock {\em Chaos, Solitons, and Fractals}, 4:5 -- 39, 1994.

\bibitem{Agha15a}
C.~Aghamohammadi and J.~P. Crutchfield.
\newblock Beyond the typical set: {Fluctuations} in intrinsic computation.
\newblock in preparation.

\bibitem{Buck90a}
J.~A. Bucklew.
\newblock {\em Large Deviation Techniques in Decision, Simulation, and
  Estimation}.
\newblock Wiley-Interscience, New York, 1990.

\bibitem{Touc09a}
H.~Touchette.
\newblock The large deviation approach to statistical mechanics.
\newblock {\em Physics Reports}, 478:1--69, 2009.

\bibitem{Ruel78}
D.~Ruelle.
\newblock {\em Thermodynamic Formalism}.
\newblock Addison-Wesley, Reading, 1978.

\bibitem{Leco07a}
V.~Lecomte, C.~Appert-Rolland, and F.~van Wijland.
\newblock Thermodynamic formalism for systems with {Markov} dynamics.
\newblock {\em J. Stat. Phys.}, 127:51--106, 2007.

\bibitem{Demb09}
A.~Dembo and O.~Zeitouni.
\newblock {\em Large deviations techniques and applications}, volume~38.
\newblock Springer Science and Business Media, 2009.

\bibitem{Crut13a}
J.~P. Crutchfield, P.~Riechers, and C.~J. Ellison.
\newblock Exact complexity: {Spectral} decomposition of intrinsic computation.
\newblock {\em Phys. Lett. A}, 380(9-10):998--1002, 2015.

\bibitem{Note2}
There are alternative statistics to which one can appeal, such as
  superstatistics \cite {Beck03a,Hane11a}. However, addressing this would take
  us too far afield at this introductory stage.

\bibitem{Bowe75a}
R.~Bowen.
\newblock {\em Equilibrium States and the Ergodic Theory of Anosov
  Diffeomorphisms}, volume 470 of {\em Lecture Notes in Mathematics}.
\newblock Springer-Verlag, Berlin, 1975.

\bibitem{Boyd16c}
A.~B. Boyd, D.~Mandal, and J.~P. Crutchfield.
\newblock Correlation-powered information engines and the thermodynamics of
  self-correction.
\newblock 2016.
\newblock arxiv.org:1606.08506 [cond-mat.stat- mech].

\bibitem{Crut99d}
J.~P. Crutchfield and P.~K. Schuster.
\newblock {\em Evolutionary Dynamics---Exploring the Interplay of Selection,
  Neutrality, Accident, and Function}.
\newblock Santa Fe Institute Series in the Sciences of Complexity. Oxford
  University Press, 2003.

\bibitem{Camp65a}
L.~L. Campbell.
\newblock A coding theorem and {Renyi's} entropy.
\newblock {\em Info. Control}, 8:423, 1965.

\bibitem{Beck03a}
C.~Beck and E.~G.~D. Cohen.
\newblock Superstatistics.
\newblock {\em Physica A}, 322:267--275, 2003.

\bibitem{Hane11a}
R.~Hanel, S.~Thurner, and M.~Gell-Mann.
\newblock Generalized entropies and the transformation group of
  superstatistics.
\newblock {\em Proc. Natl. Acad. Sci. USA}, 108(16):6390--6394, 2003.

\end{thebibliography}
